\documentclass[fleqn,twoside]{elsart}
\usepackage[dvips]{epsfig}
\usepackage{amsmath,float}
\newcommand{\figura}[3]{\epsfig{file=./#1.eps,width=#2\linewidth,angle=#3}}

\begin{document}
\begin{frontmatter}

\pagenumbering{arabic}

\title{A device to characterize  optical fibres}
\author{F.~Bosi,}
\author{S.~Burdin,}
\author{V.~Cavasinni,}
\author{D.~Costanzo,}
\author{T.~Del~Prete,}
\author{V.~Flaminio,}
\author{E.~Mazzoni,}
\author{C.~Roda,}
\author{G.~Usai\corauthref{cor}},
\corauth[cor]{Corresponding author Giulio Usai INFN Sezione di Pisa  Via Livornese 1291, 56010 S. Piero a Grado (PI)
 Tel. +39 - 050 880 438, Fax. +39 - 050 880 317}
\ead{giulio.usai@pi.infn.it}
\author{A.~Vasiljev}
\address{Dipartimento di Fisica, Universit\`{a} di Pisa and Istituto 
 Nazionale di Fisica Nucleare, Sezione di Pisa, Italy}

\begin{abstract}
ATLAS is a general purpose experiment approved for the LHC collider at CERN.
An  important component of the detector is  the central hadronic calorimeter;
for its construction more than 600,000 Wave Length Shifting (WLS) 
fibres (corresponding to a total length of  1,120~Km)  have been used.
  We have built and put into operation
 a dedicated instrument for the measurement of light yield and
attenuation length  over groups of 20~fibres at a time.
  The overall accuracy achieved  in the measurement of light yield
  (attenuation length) is   1.5\% (3\%).
 We also  report   the results obtained using this method in  the quality 
control  of a large sample of fibres.

\end{abstract}


\end{frontmatter}

\section{Introduction}

Optical fibres have been widely used in  recent years in 
High Energy Physics. Their popularity is due to several factors:  cost;
  efficiency in light transport;
  easiness to machine and adapt to various geometries.
Both scintillating fibres and WLS fibres have found
many applications in HEP detectors, in the field of 
tracking:  UA2~\cite{ua2}, $D\emptyset$~\cite{d0}, as active
 elements in  calorimeters:
 KLOE(DA$\Phi$NE)~\cite{kloe}, CHORUS~\cite{chorus}, H1(DESY)~\cite{h1},
 and as efficient 
devices to transport light from scintillators to the photomultipliers:
ZEUS(HERA) (presampler calorimeter)~\cite{zeus}, 
DELPHI(LEP)(stic detector)~\cite{delphi}.

The device presented in this paper was developed during
 the construction of the 
hadron calorimeter (Tile Calorimeter) of ATLAS~\cite{atlas}, one of the
 two major detectors 
that will acquire data in the forthcoming years at LHC. 

The Tile Calorimeter~\cite{TileCAlTDR} is an iron-scintillator calorimeter 
whose geometry has been designed
to minimize dead spaces and optimize   hermeticity.  
The two main characteristics are: the orientation of the scintillator tiles
along   the direction of the  incoming particles, and the use of 
WLS fibres to transport light to photomultipliers  (PMTs). 
The ensemble of these two choices
allowed an extremely compact design of the calorimeter with virtually no
dead spaces.

The constraints imposed by this design on the performance of fibres, 
 are severe, as we will discuss in the following section. 
 The quality of the 
fibres had to be carefully monitored all   along the production period.
The  fibres  were produced  in batches  
and the control of quality~(QC), had to be precise and fast enough to detect
and correct in real time possible deviations  from the optimal
requirements of the fibre characteristics.

\section{Requirements on the fibres}

The optical contact of scintillating tiles with WLS fibres is 
obtained by gently pressing the fibres for a fraction of their length
 against the  side of the scintillators.
 A diffuser surrounds the fibre and the scintillator,
increasing the light collection efficiency.
 Only a small fraction of light is captured by the
fibres, as  
the wavelength shifting efficiency and the effective light
attenuation length in the  fibre,
 reduce considerably the number of photons which reach the 
PMT. The overall light budget is about 40-60 photoelectrons (phe) per
 GeV deposited in a calorimeter cell.
To obtain a higher light yield, we decided to use double-clad fibres which,
 with respect to single clad fibres, ensure  a larger capture
 efficiency of light rays and longer attenuation length.
Severe requirements were placed on the wavelength shifting efficiency 
and on the effective attenuation length of light in the fibre.
Conventionally we characterize the fibres by two quantities:
\begin{enumerate}
\item the attenuation length ($\lambda$), that is obtained by measuring
 the light yield at one end, after exciting the fibre at several
positions along its length; 
\item the  light yield  ($n_0$) obtained by exciting
the fibre   at 140 cm from the end. 
\end{enumerate}

The   QC requires   these two quantities to be as large as possible
and at the same time  the spread among individual measurements to be
as small as possible.
In fact  any
large fluctuations of these quantities would produce dis-homogeneities
in the calorimeter response, resulting in  non-statistical terms
in the calorimeter resolution (the so called ``constant terms'').

The  limits on these parameters and on fluctuations in their values 
have been assesed using careful MonteCarlo simulations requiring  
 that the calorimeter resolution 
would  not be significantly  affected by fibre performance even in the long
run, when we expect  small aging effects to show up.

\section{The fibrometer}
The total length of fibres needed for the Tile Calorimeteris 1,120~Km, a  quantity 
which can be produced by a commercial firm in about one year. 
The length of  
individual fibres ranges from 73~to~230~cm and the total number of fibres 
is about 640,000.  It was not possible to measure all
the fibres, neither was it  appropriate, since during the measurement process
the fibres might  undergo damage.

Each batch of fibres was composed of about 30~production units
 called ``preforms''.
Because of the production process, the fibres within a preform are quite
uniform. Preforms could  slight differ from each other, because 
of different chemical composition, temperature, tension in pulling
 fibres, etc. 
One preform consist of about 2,000~fibres. 

We decided to measure  18~fibres of each preform, randomly chosen. 
This number allows a precise enough determination of   the mean and
 R.M.S. of  the whole preform. 
Still, more than 3,000~fibres  had to be measured.
An instrument to perform a rapid and  precise   measurement 
was therefore  designed.
We  developed a dedicated apparatus which allows a fast and precise 
 characterization of 20~fibres in parallel.

The main idea is to hold the fibres under measurement on a precision
plate which is fixed to an optical test bench. The fibres are excited 
by a movable light source that can be positioned with an x-y~table over
the fibre plate. The fibres are read by a single PMT, connected to
the fibres under measurement through  clear fibres.
Hence the only movable part of the system is the light source.

The light source is a blue  LED.
 Great care is given to the stability of the
light source and to the gain of the PMT. Both are monitored electronically;
but the stability of the system is finally checked by comparison with 
two reference fibres which have been mounted on the 
plate  at the beginning of the measurements and never moved.  

\subsection{Construction Details}

\begin{figure}[e]
  \begin{center}
    \mbox{\figura{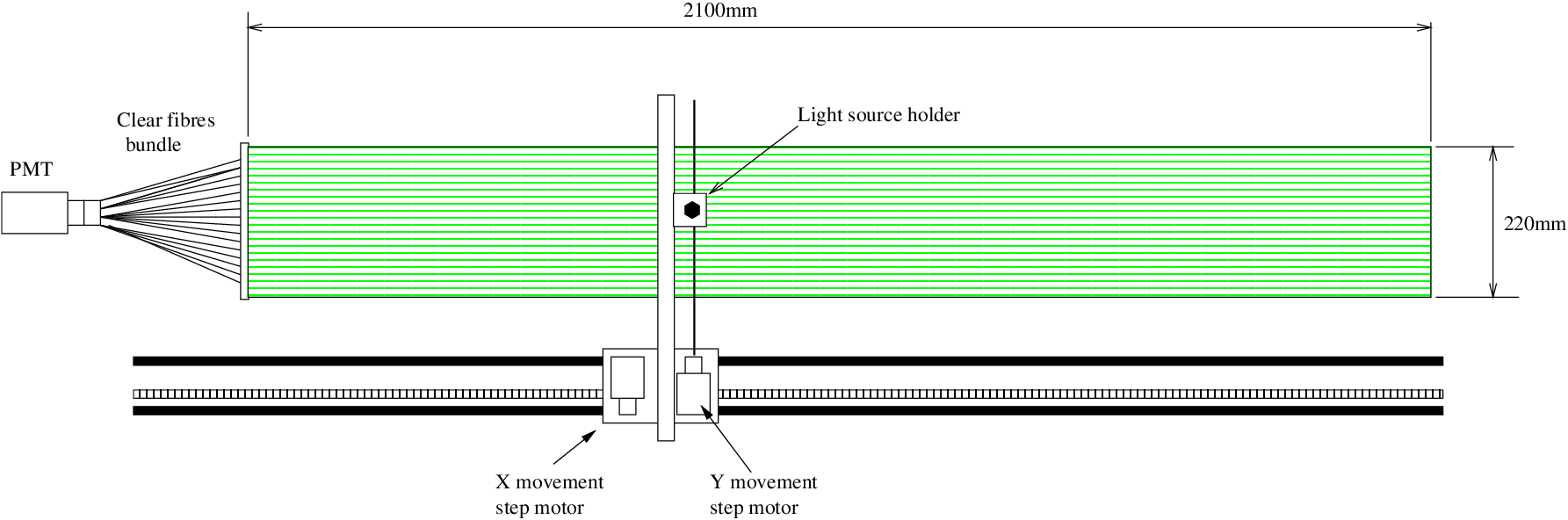}{1.15}{0}}
  \end{center}
  \caption{\sl\it General  view of the  setup for  fibre characterization.}
  \label{fig:QCsetup}
\end{figure} 

\begin{figure}[hbt]
  \begin{minipage}{.495\linewidth}
    \begin{center}
      \mbox{\figura{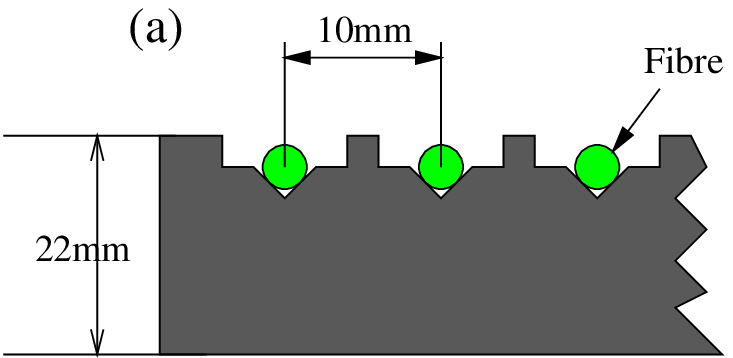}{0.9}{0}}
    \end{center}
  \end{minipage}\hfill
  \begin{minipage}{.495\linewidth}
    \begin{center}
      \mbox{\figura{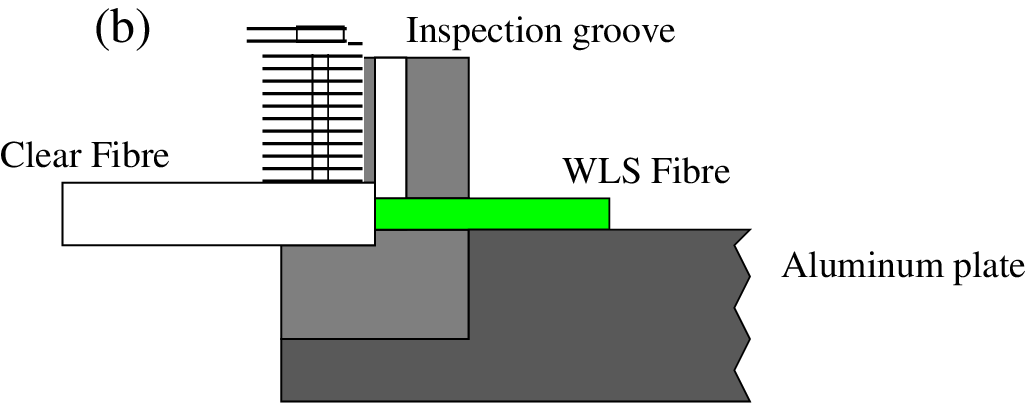}{1.2}{0}}
    \end{center}
  \end{minipage}\hfill
  \caption{\sl\small  Side  view of fibre plate (a) and  detail of coupling
between clear and WLS fibres (b).}
  \protect\label{fig:plate_multifi}
\end{figure}

A sketch of the fibrometer is shown in Fig.~\ref{fig:QCsetup}.
The aluminum plate which holds in place the fibres is 2~m long  and is
installed on a granite  optical table.
 The aluminum plate was machined 
to obtain 20~parallel V-shaped grooves that  house the fibres 
under measurement. The distance between adjacent  grooves is 1~cm,
 each groove being 0.65~mm deep with an  angle of 90~degrees. 
Details of the grooves geometry are shown in
Fig.~\ref{fig:plate_multifi} (a) 
which shows a 
cross section of the plate. A  4~mm high wall between the grooves
reduces any possible  cross-talk between nearby fibres.
The plate is black anodized to avoid spurious reflections.
Fibres are kept fixed to the plate at three positions: in the middle and at
the two ends. At these points the  shape of the grooves is
appropriately modified in such a way that the fibres can be hold in
place by the preassure applied through  the use of small  soft-rubber
pieces.
 Table and fibrometer are placed inside a light-tight dark room.

The fibres are excited using a light source with an emission spectrum
similar to the one of the Tile Calorimeter plastic scintillators.
The absolute response of fibres to this light source is not identical 
to the scintillation light, since the spectral emission is not exactly
the same. 
This, however, does not affect the results of the QC.
As a light source we have used a very intense 
blue~LED~($\bar{\lambda}$=430~nm)\footnote{LEDTRONICS type  BP280CWB1K.}~DC operated.
 The LED was mounted on a x-y~table movable in 
precise steps (3~mm along  and 5~$\mu$m across the fibres, for each
clock pulse).
The LED was kept within a PVC case at about 20~cm from the fibres.
 The light
is collimated with a 0.7~mm slit, placed at 5~mm from the fibres.
 The size of the light spot
was 0.8~mm at the fibre.
 Cross-talk effects between fibres were, under these
conditions, negligible.

The LED source is clearly much easier and safer to handle than a 
light source made of a plastic scintillator and a radioactive source.
The latter is much more stable and would reproduce exactly the 
light spectrum of the detector. The LED on the contrary has to be calibrated
with respect to the scintillator light and continuously monitored
in intensity. After having considered the pros and cons of the two
solutions the LED system was finally chosen.

 Fig.~\ref{fig:QCsetup} shows some details 
of the x-y~table used for the positioning of the light source.
Two rails, mounted rigidly on an optical bench, and a precise gear 
allow the movement of a motorized head along the fibres (x-direction).
The head supports a light aluminum arm, orthogonal to the rails,
on which the light source is mounted. Precision rails guide the 
movement across the fibres (y-direction) of the light source.
Both movements are driven by step motors\footnote{produced by Sigma Instruments Inc. Model 202223D200-F6.}
under computer control. No sensor was used to monitor the position
of the light source, which is obtained by counting the number of steps of
the motors. 
This implies that any clock pulse lost in the motor driving
electronics would result in a wrong positioning of the light source.
From the following discussion it will be clear that the system is
self-calibrating in position across the fibres where the measurement steps are
very fine (0.1~mm).
A large step size~(10~cm), between consecutive measurements, was
chosen for the movement along the fibres.
Therefore the loss of one or a  few  clock pulses would be unimportant.

\begin{figure}[e]
  \begin{center}
    \mbox{\figura{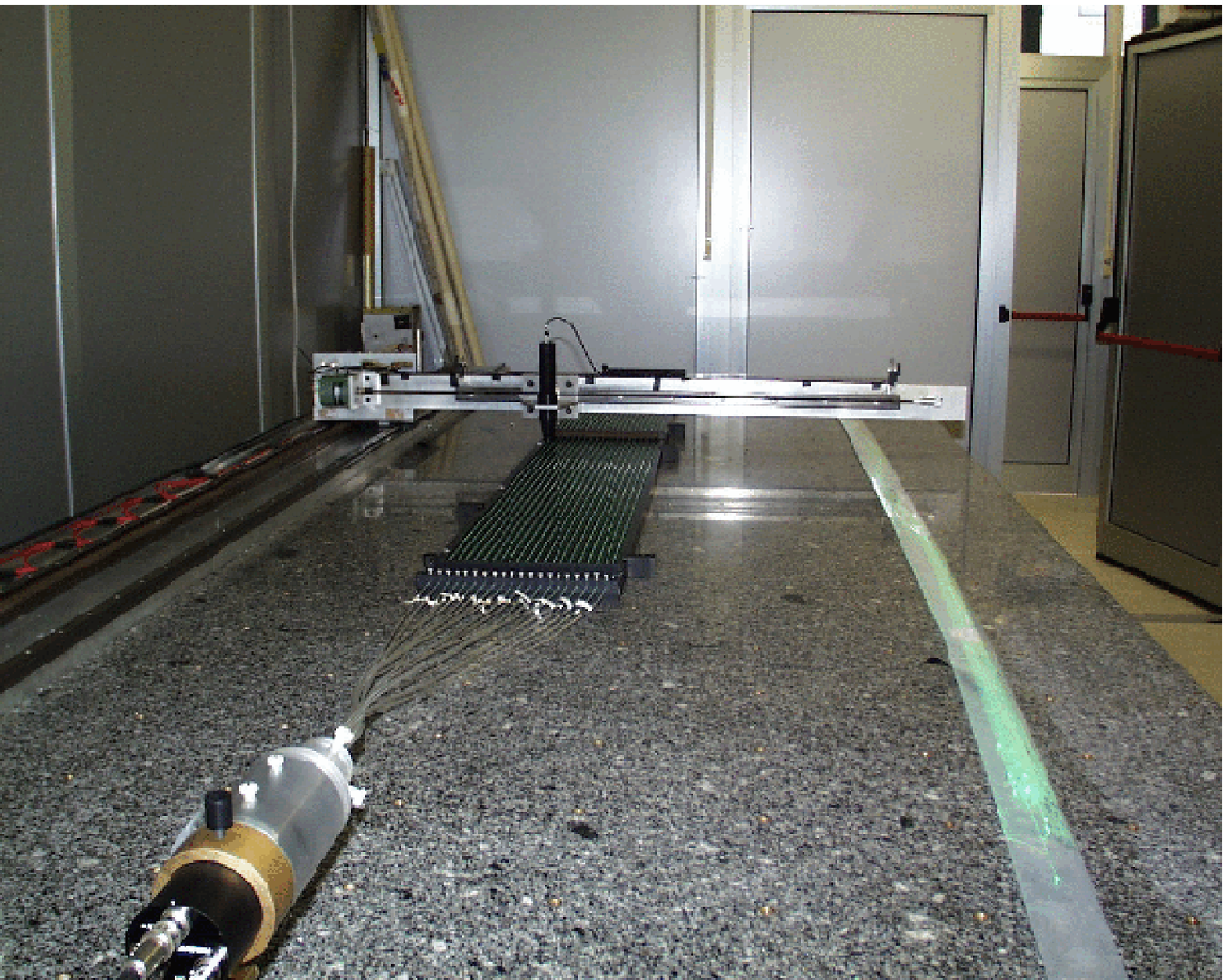}{0.8}{0}}
  \end{center}
  \caption{\sl\it Photograph of  the  setup for  fibre characterization.}
  \label{fig:photo_multifi}
\end{figure}

Twenty, 2~mm diameter, clear fibres provide the light
collection\footnote{ Pol.Hi.Tech. fibres type OP. }.
One end of the clear fibre is optically machined, mounted on
the fibre plate and held in position by a teflon screw 
(Fig.~\ref{fig:plate_multifi}~b).
The fibres under measurement are optically coupled to the clear ones
by  air coupling.
The other ends of the clear fibres are glued together inside a 
Plexiglas tube and optically machined. The clear fibre bundle faces
a PMT\footnote{ Hamamatsu H3178 $\phi$=38~mm.}
through a light mixer. A single PMT receives the light of all the 
twenty fibres which are excited one at a time.
The signals from the PMT are sampled every 30~ms by a digital 
multi-meter\footnote{ Keithley Mod. 2700.}.
Fig.~\ref{fig:photo_multifi}  shows the granite table with the fibrometer 
mounted on it.

\begin{figure}[e]
  \begin{center}
    \mbox{\figura{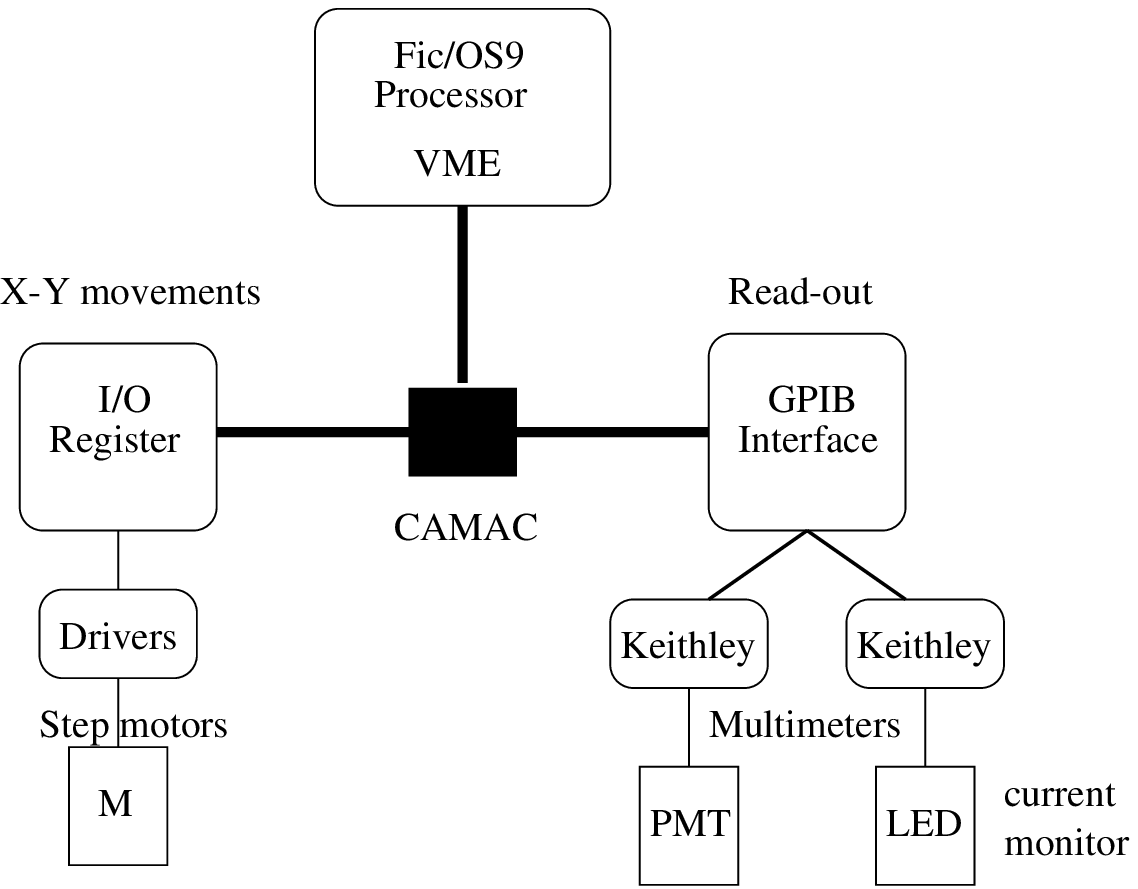}{0.5}{0}}
  \end{center}
  \caption{\sl\it Schematics of the readout electronics. }
\protect\label{fig:elect_sch}
\end{figure}

The Data Acquisition System controls the operation of  the 
stepping motors and performs the synchronous readout of the 
 multi-meter.
It is based on a FIC OS9\footnote{Creative Eletronic System mod. FIC 8232.}
housed in a VME crate. The I/O electronics is CAMAC based, a VME-CAMAC
interface\footnote{Creative Eletronic System Mod. CBD8210.}
provides the interconnections between the two systems.
Fig.~\ref{fig:elect_sch} shows a schematics of the readout electronics.

An important point in the construction of the fibrometer is the
stability in parallelism between the plate that holds the fibres and the light source.
Any deviation from parallelism during the movement of the source would
result in a change of the source-fibre distance, thus affecting the
determination of fibre parameters.
There are two sources of error of this type.
The first is almost unavoidable in our system and is due to the 
elasticity of the fibres. The fibres would not perfectly adhere to the
grooves but rather twist a bit. We have measured a maximum of +0.2~mm
between the nominal and actual position of the fibre.
The second source of errors is the mechanical tolerances in the 
construction of the x-y table and its
 positioning on the optical test bench.  
The use of a granite optical table as a support for the fibrometer  greatly
reduced the latter problem. The  precise mechanical construction of
the plate  reduces the 
effect of non-parallelism to the level of 0.2~mm.
This uncertainty would  still be  too large with a source  positioned 
near (few mm)  the fibre. For this reason we decided to keep
the light source at 20~cm from the plate. This has two advantages:
it strongly reduces the effect of non-parallelism  and  
yields  a very narrow beam of light (smaller than  the fibre diameter).

\begin{figure}[e]
  \begin{center}
    \mbox{\figura{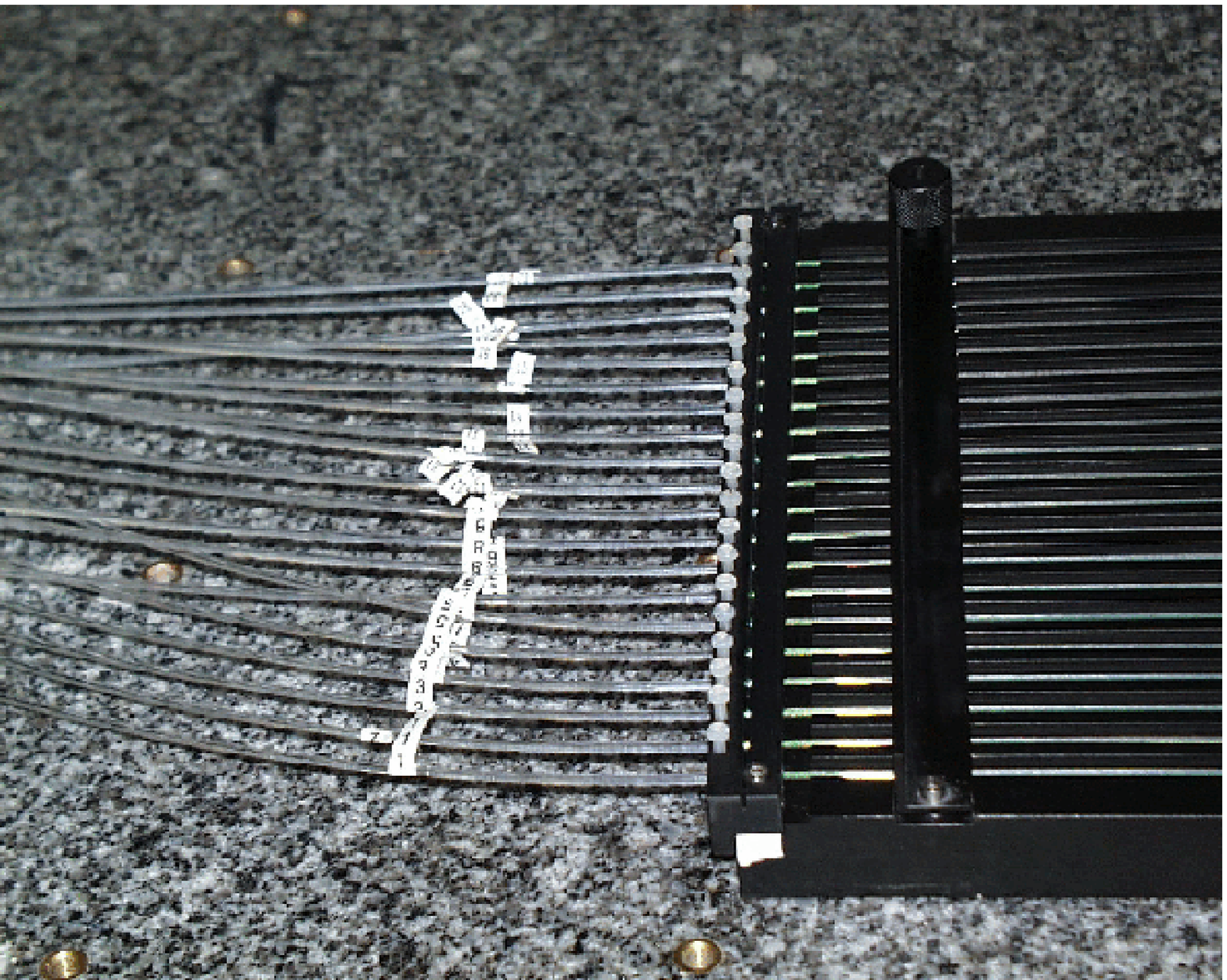}{0.7}{0}}
  \end{center}
  \caption{\sl\it Setup used for coupling  clear and WLS fibres.}
\protect\label{fig:coupling}
\end{figure} 
The second important point is the optical coupling of the test fibres to the
clear fibres. 
The clear and test fibre ends which are in optical contact
are diamond cut and polished. This is easily obtained  by a dedicated 
tool\footnote{AVTECH Inc. Schneider drive, 625 South Elim, Illinois.}.
Since no optical interface (grease or optical coupler) is placed between 
the fibres, their relative geometry must be well reproducible.
Fig's.~\ref{fig:plate_multifi}~(b) and~\ref{fig:coupling} show how 
the coupling between the fibres has been 
obtained in our system.
A carefully machined PVC piece is positioned 
at the end of the fibre plate. On it, precision holes guide the fibres,
clear and WLS,
which are kept coaxial till they touch. Clear fibres are held in place by
Teflon screws, test fibres are independently held against the plate.
The fibre positions can be visually inspected through a groove.

\section{Fibre measurement}

The first step in the measurement process is the fibre preparation.
 The diamond cut and polishing of the fibre end facing the clear 
fibre has alredy been described. 
The other end of the fibre should not reflect any light. 
This is obtained by a complete blackening
of the fibre end opposite  the readout side. 
The tip of the fibre is cut roughly with scissors at 45~degrees and 
painted with black oil ink.

The fibres are then numbered, for later reference, placed on the
plate and positioned in contact with the clear fibres. We found 
useful to use a vacuum cleaner to remove residual dust while 
sliding the fibres in place.

The measurements were then performed in a completely automatic way, 
under computer control.

After turning on the system  we wait for 30~minutes before data taking
in order to warm up the electronics and the PMT.
 Then  we start
the raster scan of the fibre plate. This is done in coarse steps long  the 
fibre direction (x-axis) and in precise steps across the fibres
(y-axis).
\begin{itemize}

\item The LED, originally in a fixed starting position (home) 
is moved to a point 70~cm away from the clear-WLS fibre contact (here
the contribution of light escaping from the fibre core and undergoing
total reflection at the clad-air boundary is alredy negligibile)
and a scan is performed along y in 100$\mu$m steps.
At each step a record is taken  of the PMT and of the LED currents; 
\item Subsequently a new scan is performed along y at 10~cm from the
first (80~cm from the clear-WLS fibre contact) and the PMT and LED
currents are  again recorded at each step.
 The procedure is repeated 12 times, till the end of fibre is reached;
\item At the end of the measurement the LED  is brought back to its
initial position (homing). 
 If, as discussed above, the stepping  motors
 have lost one or a few clock pulses  the homing would not be
correctly achieved. 
In such case the measurement would be repeated.
\end{itemize}
The measurement of 20~fibres takes about 25~minutes and the data 
recorded  are then  processed off-line.
\begin{figure}[hbt]
  \begin{minipage}{.495\linewidth}
    \begin{center}
      \mbox{\figura{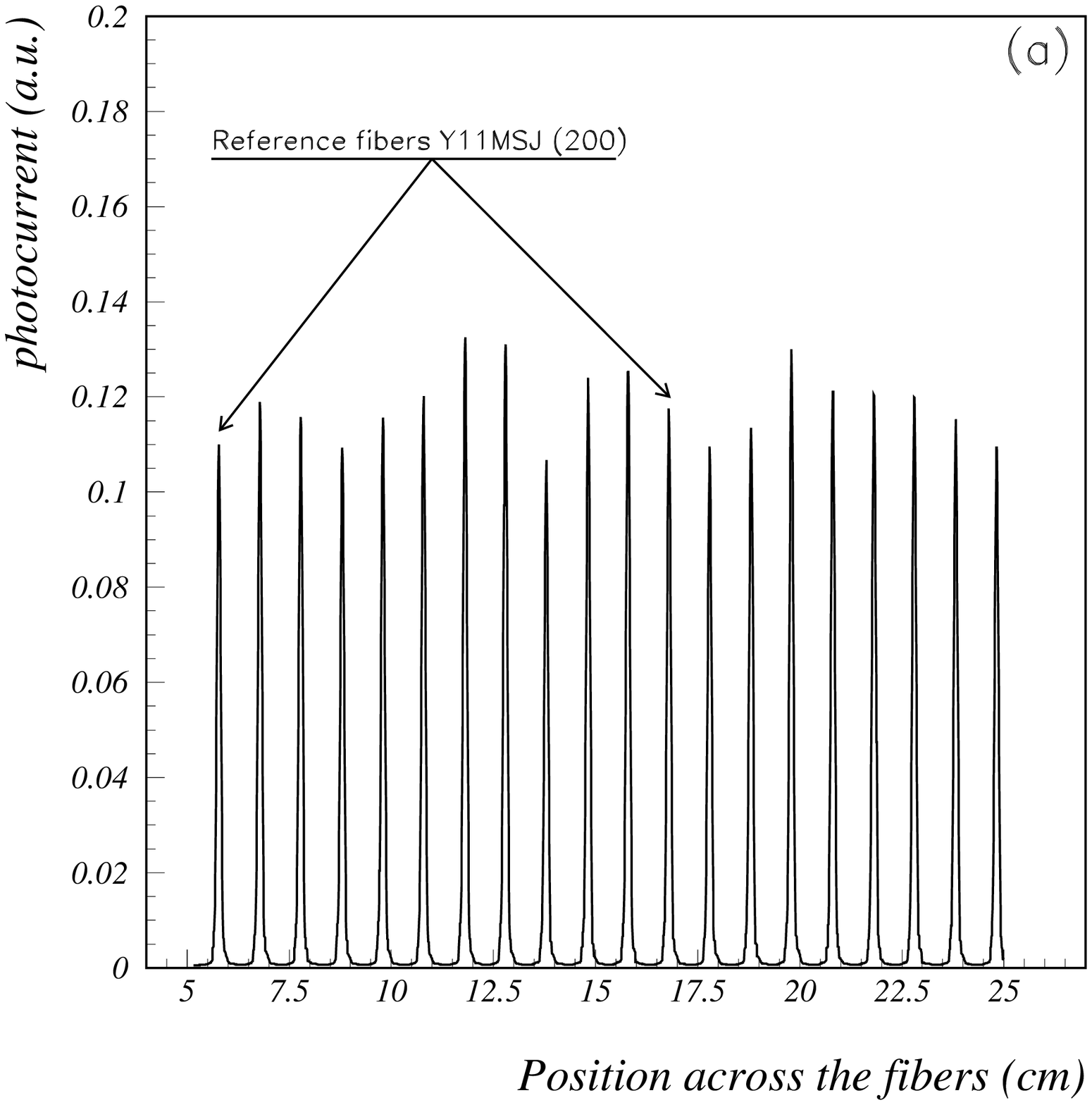}{0.9}{0}}
    \end{center}
  \end{minipage}\hfill
  \begin{minipage}{.495\linewidth}
    \begin{center}
      \mbox{\figura{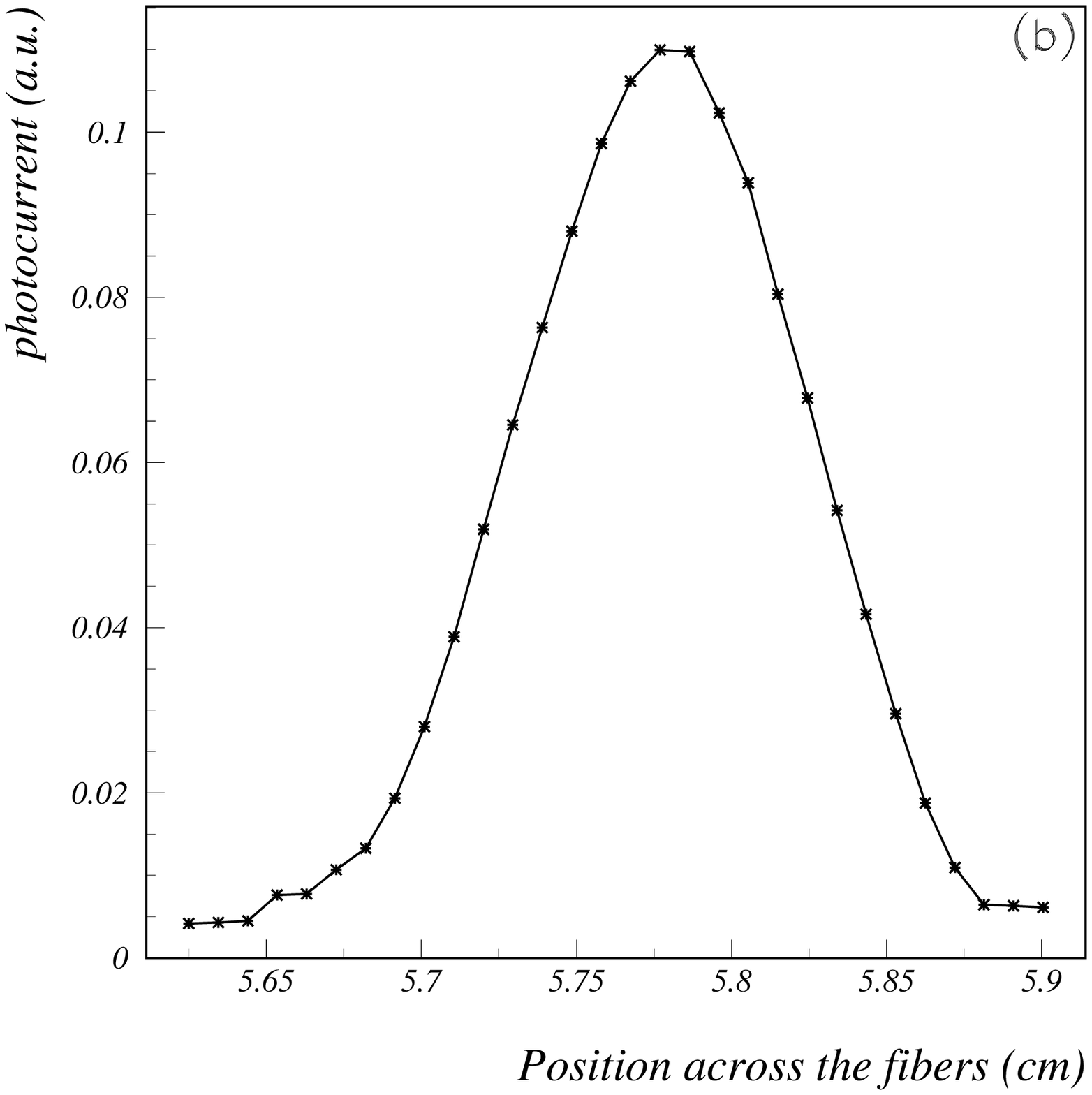}{0.9}{0}}
    \end{center}
  \end{minipage}\hfill
  \caption{\sl\small Photo current as a function of position during a
    scanning across 20 fibres (a) and detail from a scanning
    across a single fibre (b).}
  \protect\label{fig:multipeak}
\end{figure}

At a fixed x, the scan across the fibres produces peaks of 
photo current, when the LED crosses  a fibre. 
Fig.~\ref{fig:multipeak}~(a),  shows the 20~peaks
in a transverse scan corresponding to the 20~fibres under measurement.
Each fibre is sampled at several points and its position is thus very well 
defined. Fig.~\ref{fig:multipeak}~(b), shows the detail of one of the peaks.

\begin{figure}
  \begin{center}
    \begin{minipage}{.495\linewidth}
      \mbox{\figura{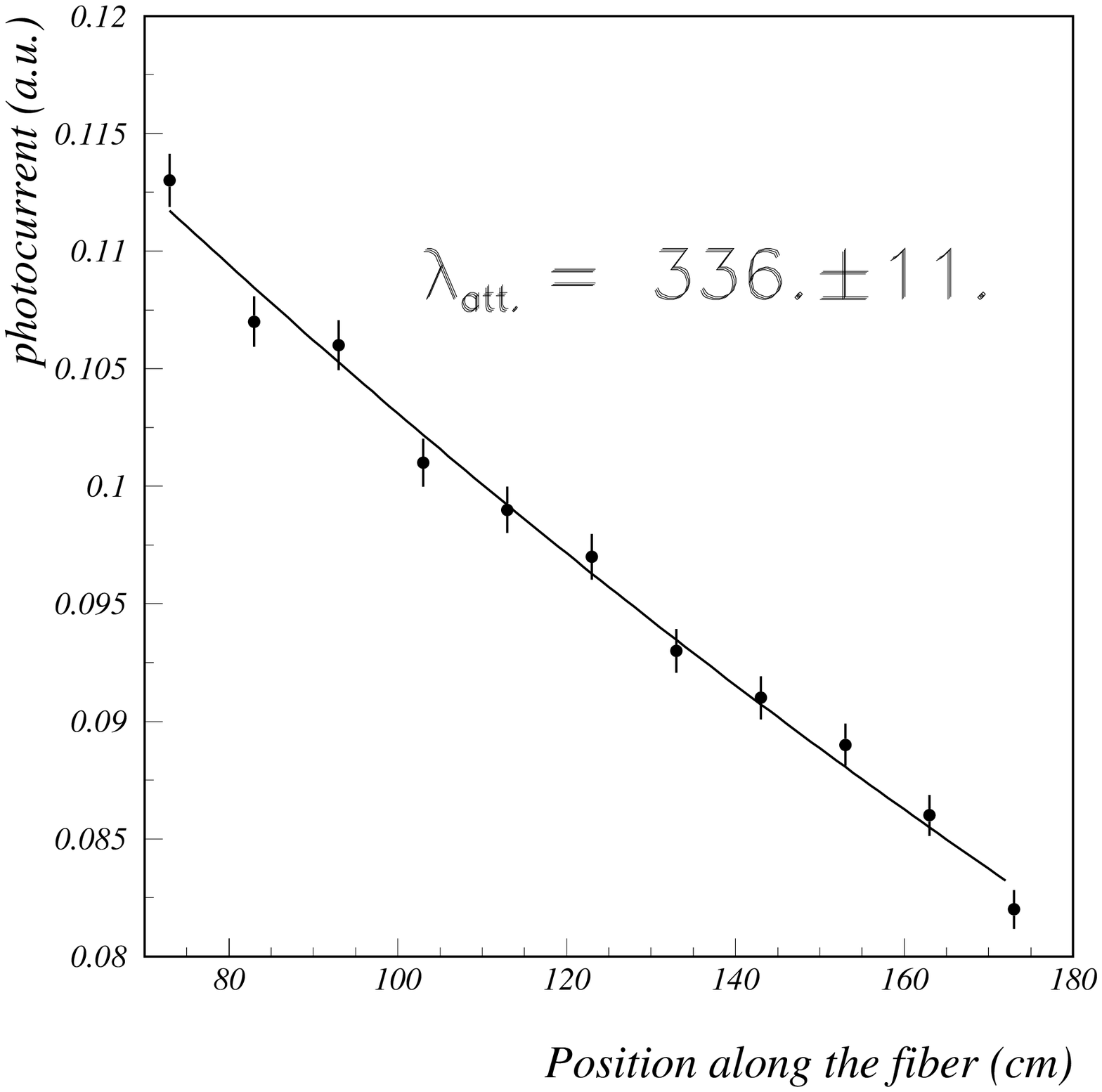}{0.9}{0}}
    \end{minipage}
  \end{center}
  \caption{\sl\it An example of attenuation curve.}
  \protect\label{fig:att_curve}
\end{figure}

The next step is to determine  the  light yield corresponding to each
of the peaks at each position.
We  have attempted the use of the following variables:
\begin{itemize}
\item  the maximum of the distribution;
\item the area under  the peak;
\item the maximum of a Gaussian fit to the peak;
\item  the area of a Gaussian fit to the peak.
\end{itemize}  
We have finally chosen the one which gave the smallest variance in 
repeated measurements. This turned out to be  the maximum of the distribution 
(ratio between R.M.S. and average about $0.9\%$). 
 
Fig.~\ref{fig:att_curve} shows the attenuation curve of a fibre,
i.e. the PMT current obtained as just described, as a function of the
LED position along the fibre.
Fitting an  exponential function to the
data we  extract
the parameters needed for the fibre characterization. 

\subsection{Channel intercalibration}

The same fibre measured after placing it  in different positions (grooves) on the plate may
give different light responses. 
The relative difference between measurements performed in  different
grooves cannot be reduced below $20\%$, mainly because of  
 differences among the different  clear fibres and their coupling to the PMT.
The different positions  have thus to be intercalibrated. 
This is performed following  a procedure which is repeated before each new batch
of fibres is delivered (2-3 months).
The intercalibration is carried out by  measuring $20$~times, as described above, 
a set of $20$~fibres.
Each measurement is done after cyclic permutation of the $20$~fibres
in the $20$~grooves.
We thus know the response of each  fibre in all the positions, which
allows  redundancy  in the  cross check of the calibration constants.
 $19$~average calibration constants are obtained as a result of this
procedure.
The intercalibration procedure also implies that the measurement of a fibre 
can be reproduced once it is removed from the plate.
We find the result to be stable  within 1\% in repeated
intercalibration procedures.

\section{Stability and precision}
Two important points are:
\begin{itemize}
\item  the precision of the measurement and  the suppression of 
systematic effects;
\item the stability of the system.
\end{itemize}

\begin{figure}[e]
 \begin{minipage}{.495\linewidth}
    \begin{center}
      \mbox{\figura{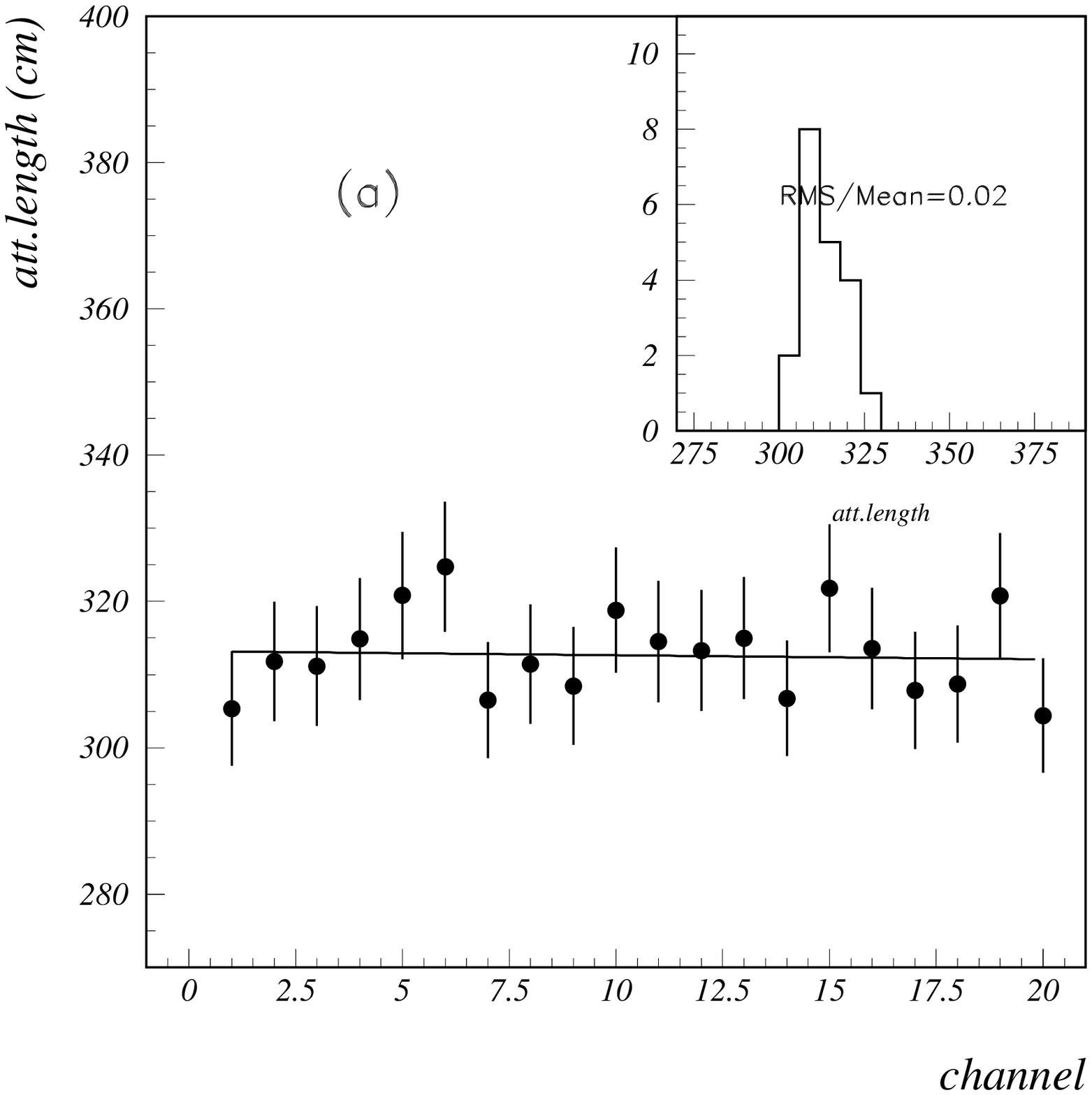}{1.1}{0}}
    \end{center}
  \end{minipage}\hfill
  \begin{minipage}{.495\linewidth}
    \begin{center}
      \mbox{\figura{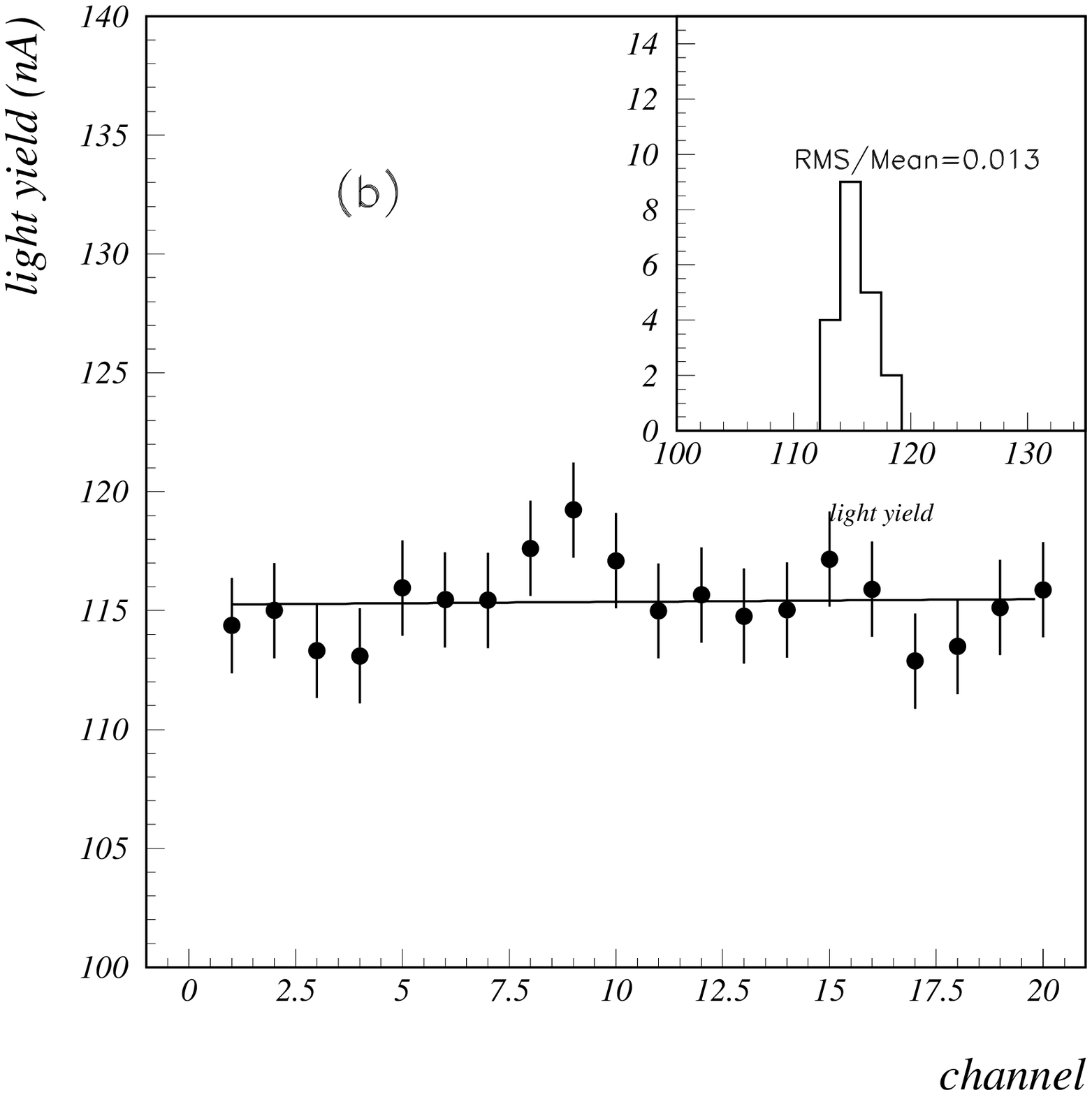}{1.1}{0}}
    \end{center}
  \end{minipage}\hfill 
  \caption{\sl\it Distributions of the attenuation length  (a)
    and  light yield (b) of a particular fibre
           measured  after subsequently placing it in each of 
the 20 grooves of the setup.}
 \protect\label{fig:dispersion}
\end{figure}
We have checked the residual systematic effects by measuring
the same fibre  after placing it in each of the $20$~positions of the plate.
If we have been able to correct for all the effects described above,
 the attenuation length and the light yield  must be independent of
the channel number. Fig.~\ref{fig:dispersion}~(a),  shows the 
attenuation length obtained for the given fibre placed in each of  the 20
grooves  of the plate.
There is no trend or other dependence on the channel number.
The  precision in the measurement of $\lambda$ is evaluated by 
the ratio $\sigma_{\lambda}/\bar{\lambda}$ which is equal to $2\%$ 
Fig.~\ref{fig:dispersion}~(b), shows the results for the light yield.
In this case as well  the light yield is independent of the channel
number.
 The  ratio $\sigma_{n_{0}}/\bar{n_0}=1.3\%$ gives   the
precision in the measurement of the light yield.

\begin{figure}[e]
 \begin{minipage}{.495\linewidth}
    \begin{center}
      \mbox{\figura{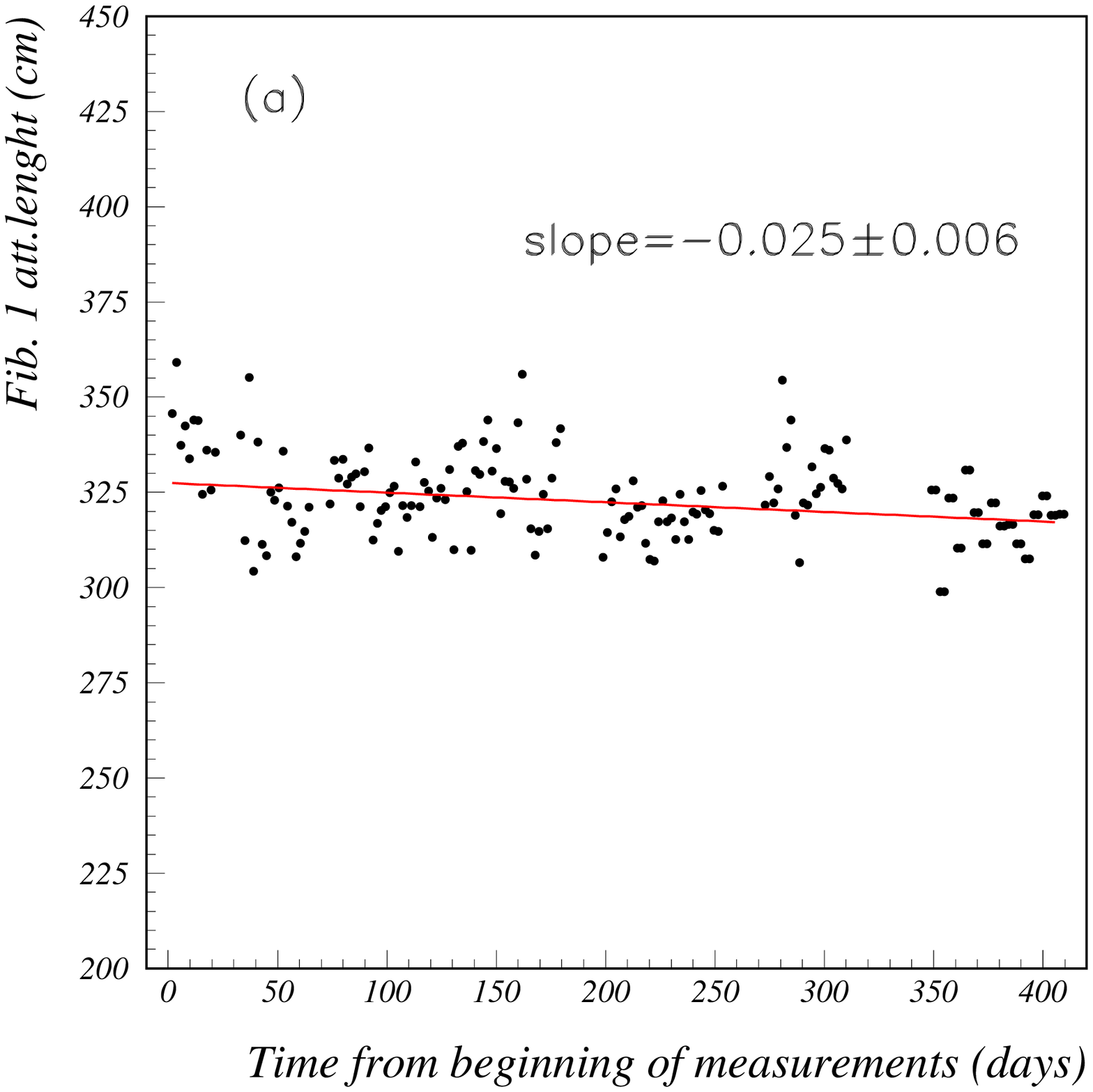}{1.1}{0}}
    \end{center}
  \end{minipage}\hfill
  \begin{minipage}{.495\linewidth}
    \begin{center}
      \mbox{\figura{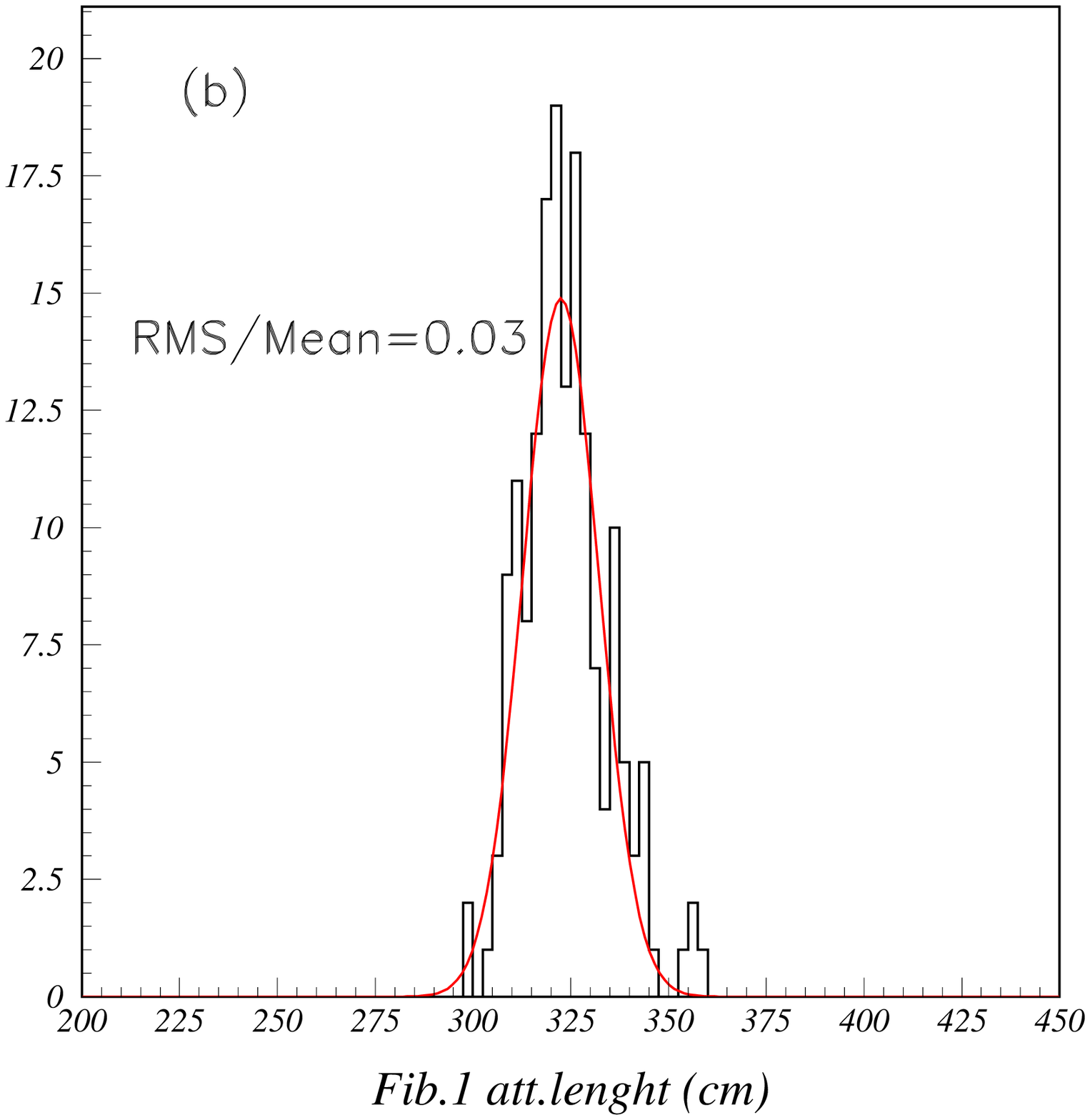}{1.1}{0}}
    \end{center}
  \end{minipage}\hfill
  \caption{\sl\it  Results of the measurements of attenuation length  of 
   reference  fibre \#1  as a function of time (a).   Histogram of
    results (b).}
\label{fig:attfib1vstime}
\end{figure}
The fibre characterization process lasted  more than one year. 
During the entire  period the  stability had to be kept
 to better than 1\%, in order to compare the absolute 
light yield of the fibres under test.
It is not easy to keep an absolute calibration at this level 
over a period of  many months. 
We thus  decided to use two fibres as reference. These were   
 kept fixed in grooves \#~1 and \#~11 
(see figure~\ref{fig:multipeak}~(a)) and untouched  
throughout the entire  period.
The reference fibres  are removed from the plate only  
 during the calibration of channels.
The light yield  measurement was always normalized  to the
first of the two fibres, while  the ratio of the light yield 
of the two reference fibres was used as a  monitor of the system stability.
We have checked that the goal of maintaining the 
stability of the measurement at the level of 1\% over one year 
was achieved.  

Fig.~\ref{fig:attfib1vstime} shows the results obtained for the 
attenuation length ($\lambda$) of reference  
fibre \#~1  as a  function of time.  
The measurement shows a  slight but significant variation   with
time (slope $-0.025 \pm 0.006$), probably due to a small  progressive
damage of the fibre.
  We   note that this effect is smaller than  the fluctuations  of
individual  measurements,  and the  overall relative
dispersion  $\sigma_{\lambda}/\bar{\lambda}= 3\%$   is compatible with
the precision of a single  measurement.
\begin{figure}[e]
 \begin{minipage}{.495\linewidth}
    \begin{center}
      \mbox{\figura{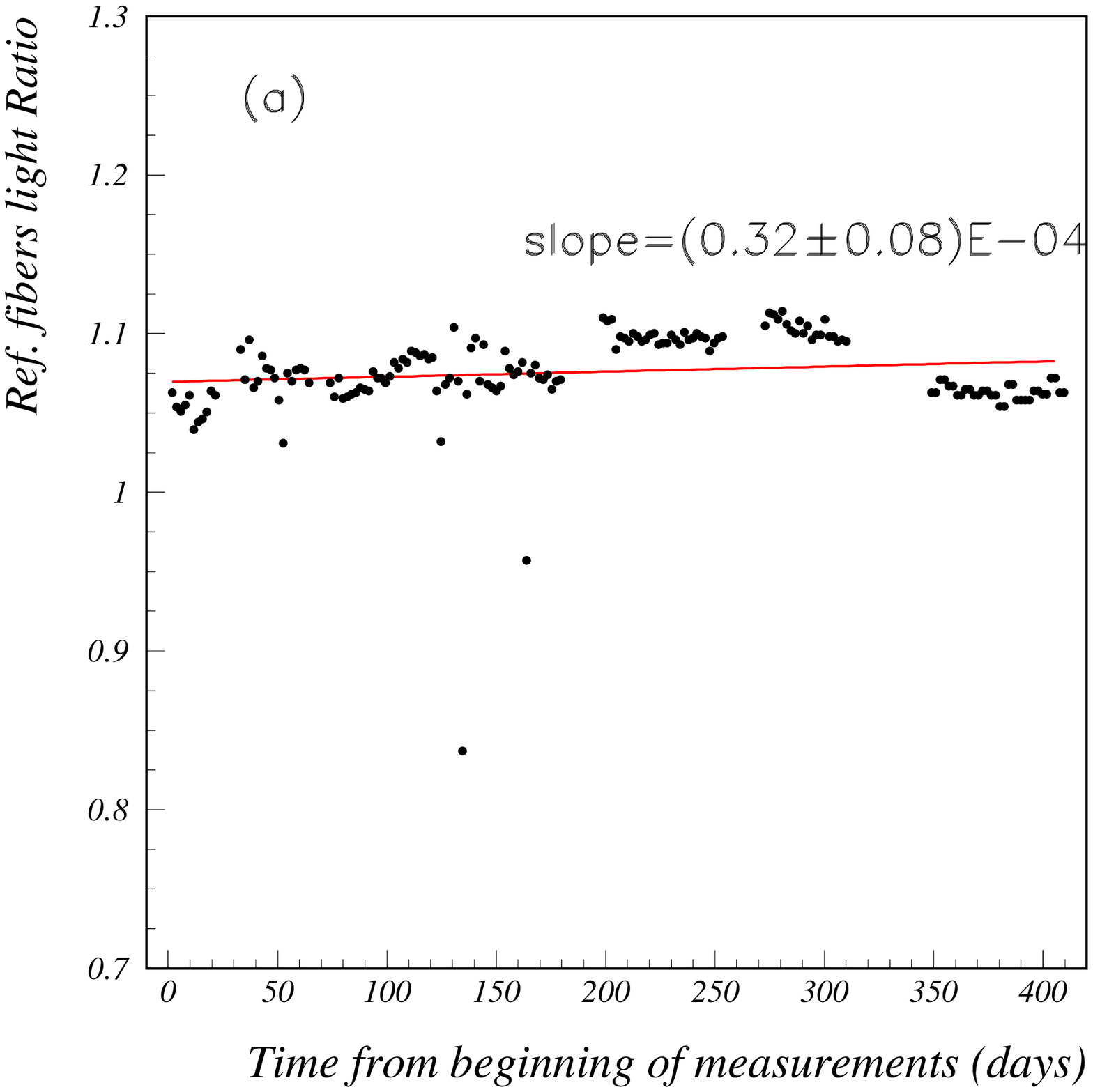}{1.1}{0}}
    \end{center}
  \end{minipage}\hfill
  \begin{minipage}{.495\linewidth}
    \begin{center}
      \mbox{\figura{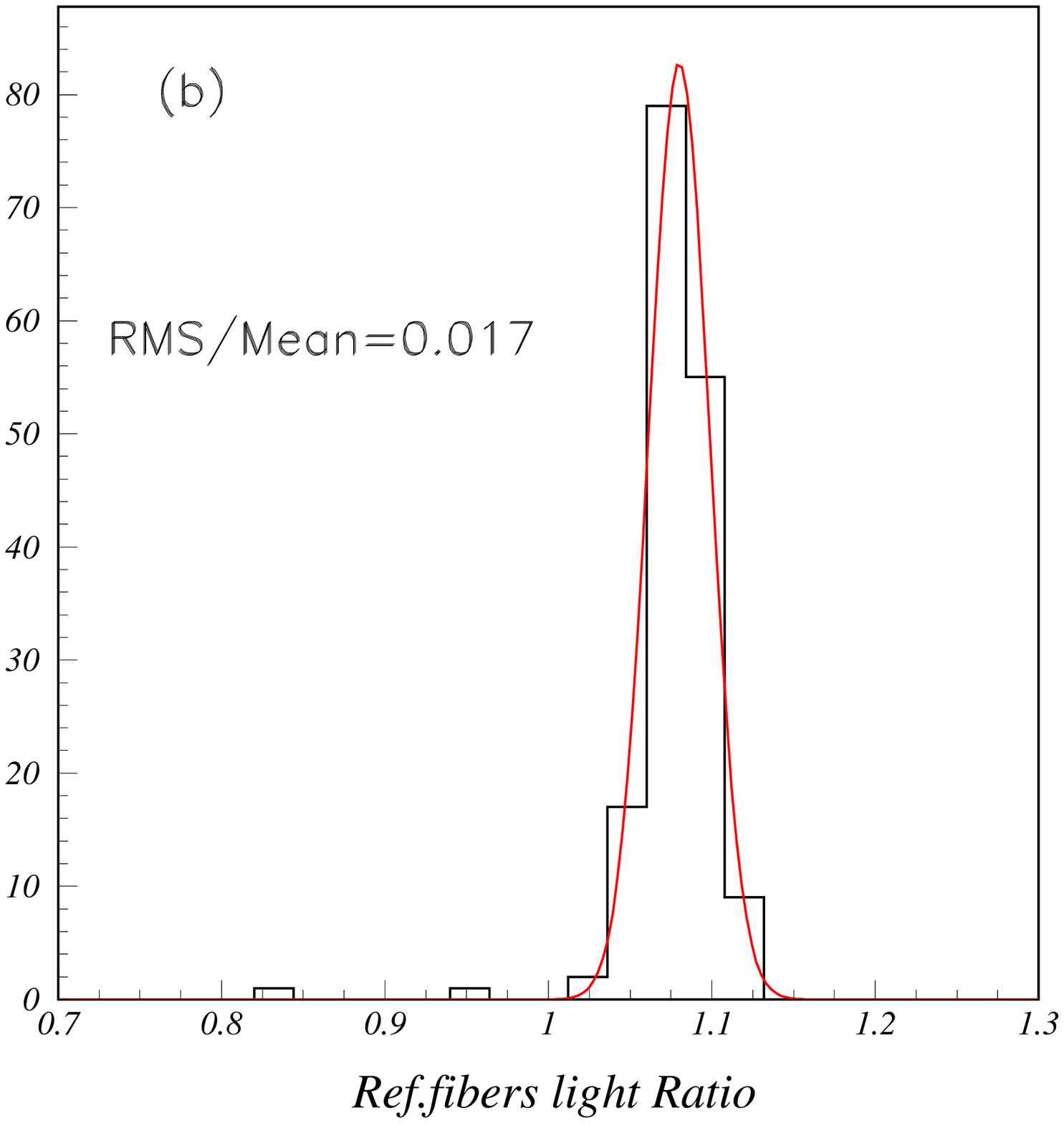}{1.1}{0}}
    \end{center}
  \end{minipage}\hfill
  \caption{\sl\it   Ratio  of the light yield of the  references
    fibres \#1 and \#11  as a function of time (a). Histogram of
    results (b).}
\label{fig:lyfib11vstime}
\end{figure}
The ratio of the light yield  of reference fibres is shown in
 Fig.~\ref{fig:lyfib11vstime} as  function of time.
It is constant and independent of 
time, with a relative dispersion: $\sigma_{n_{0}}/\bar{n_0} = 1.7\%$. 
The two anomalously low  data point seen in
 fig.~\ref{fig:lyfib11vstime}  are due to an
 incorrect positionining  of fiber \#1  on the plate caused by linear thermal
expansion.  
Since the  reference fibres were   kept fixed  on the plate at both 
ends, temperature  variations could result in a bending of the fibres
 and thus in a wrong measurement.
This  problem  has been  
solved  by  rigidly fixing  on the plate   only the  readout end
 of the fibre, while  
keeping the  other   fixed  in such  a way that it  can slide. 
The  measurements  have then  been repeated. 
The relative stability of  the measurements performed  over  
 more than a  year  turn out to be  about 1\%.  

\subsection{Systematics on attenuation length }

The measured attenuation length of a fibre is  dependent
upon the angle under which the light is collected at the fibre end.
Light collected at large angles is more likely to correspond to
photons undergoing many large-angle reflections at the 
core-cladding  interface   thus yielding  a shorter 
attenuation length.
The opposite is true of light collected at small angles, corresponding
to  photons undergoing few reflections. 
The light which reaches the end of the fibre is a mixture of all
these components, and the  attenuation length  depends on which component
is detected.
The two extreme  cases are:
\begin{itemize}
\item  light detected at a very  small angle to the fibre axis.
 Here the  attenuation length will approximate  that of the   
 core material (about 400~cm in our  case, where  the fibre core has 
polistyrene  as base material);
\item light detected at all angles:  in this
case the light yield will be larger,  but
the attenuation length  shorter (about 250~cm)\cite{method89}.
\end{itemize}   
In our apparatus the light  is collected by the PMT through a
single-clad clear fibre, and the optical rays that  are impinging on 
the clear fibre at an angle with the axis larger  than  the critical 
value~(${19.4}^{\circ})$
 escape and  cannot thus  reach  the PMT.
As a conseguence the attenuation length  that we measure is larger by
$30-40\%$ than the one where  all the light is collected.
When comparing   the measured  attenuation lengths  obtained using  
different  experimental procedures, one  has  to take into account 
 the  details of collection of the light rays
by the  PMT.
We have performed careful studies of this problem,  both
experimentally~\cite{method99}
  and through Montecarlo simulations~\cite{comment_att}.
The results show  that the
differences among different measurements agree  when the appropriate
 geometry of light collection is taken into account.

\section{Results of the QC for the Tile Calorimeter}
This device and the measurement procedure described above 
has been set up for the quality control of the  WLS fibres used by
ATLAS in the Tile hadron calorimeter. A fraction  of the fibres 
(75\%, corresponding to about 460,000 fibres) have been qualified 
in our laboratory using  
this device and following the above procedures.

\begin{figure}[e]
\begin{center}
\mbox{\figura{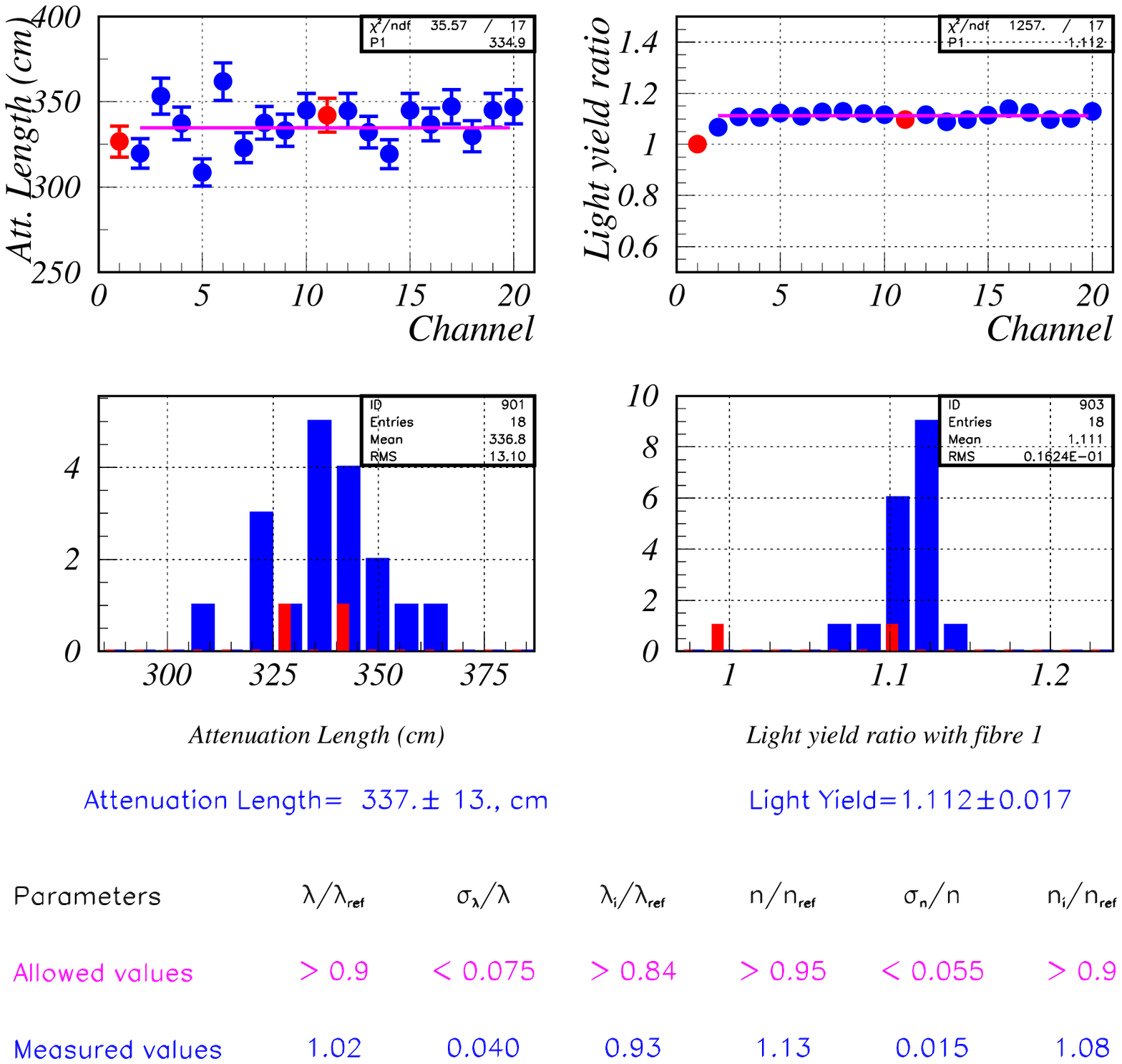}{0.7}{0}}
\end{center}
\caption{\sl\it Light yield and attenuation length measured for
18~fibres from  one of the measured preforms.}
\label{fig:prefesample}
\end{figure}

\begin{figure}[e]
 \begin{minipage}{.495\linewidth}
    \begin{center}
      \mbox{\figura{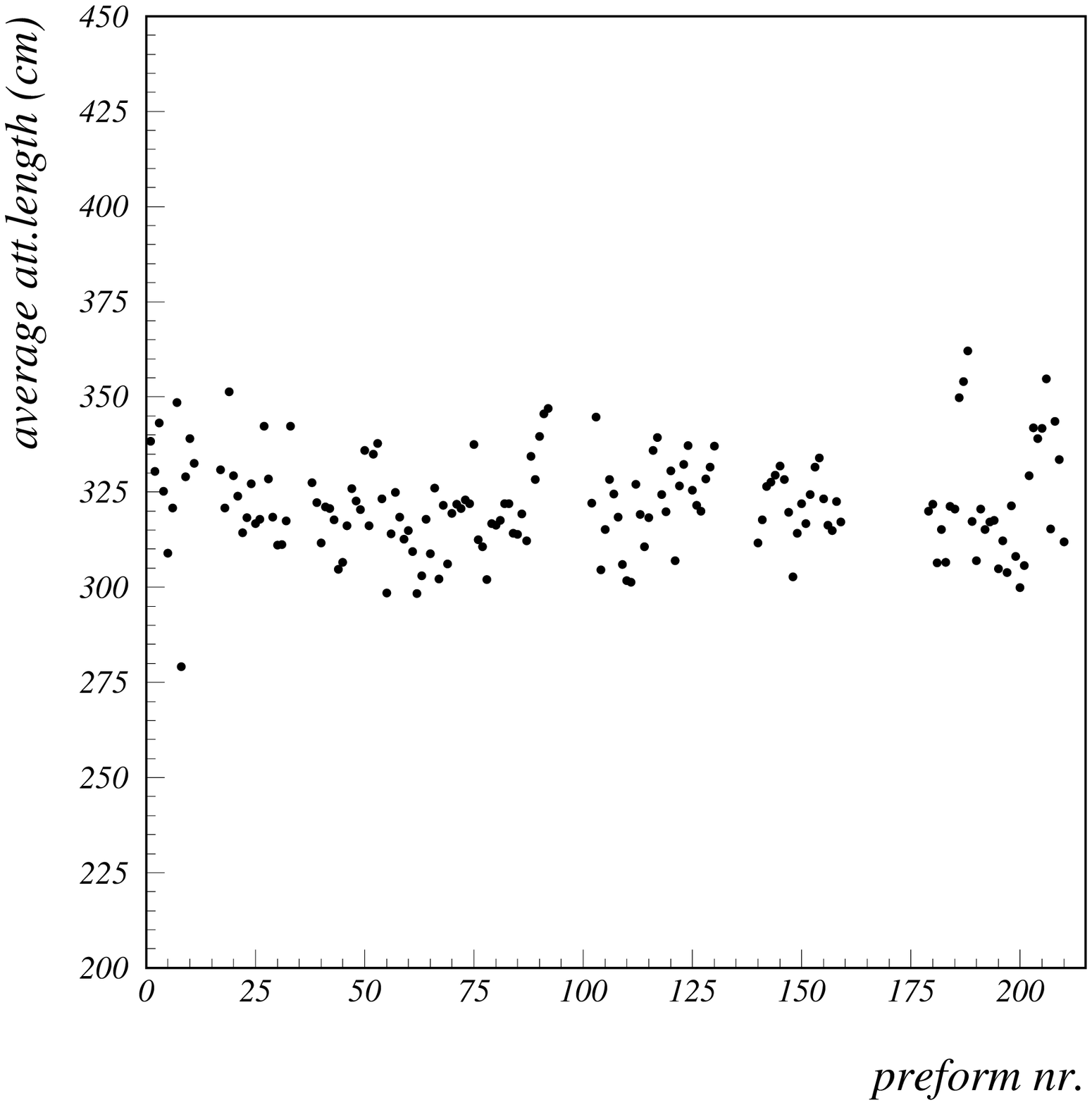}{1.1}{0}}
    \end{center}
  \end{minipage}\hfill
  \begin{minipage}{.495\linewidth}
    \begin{center}
      \mbox{\figura{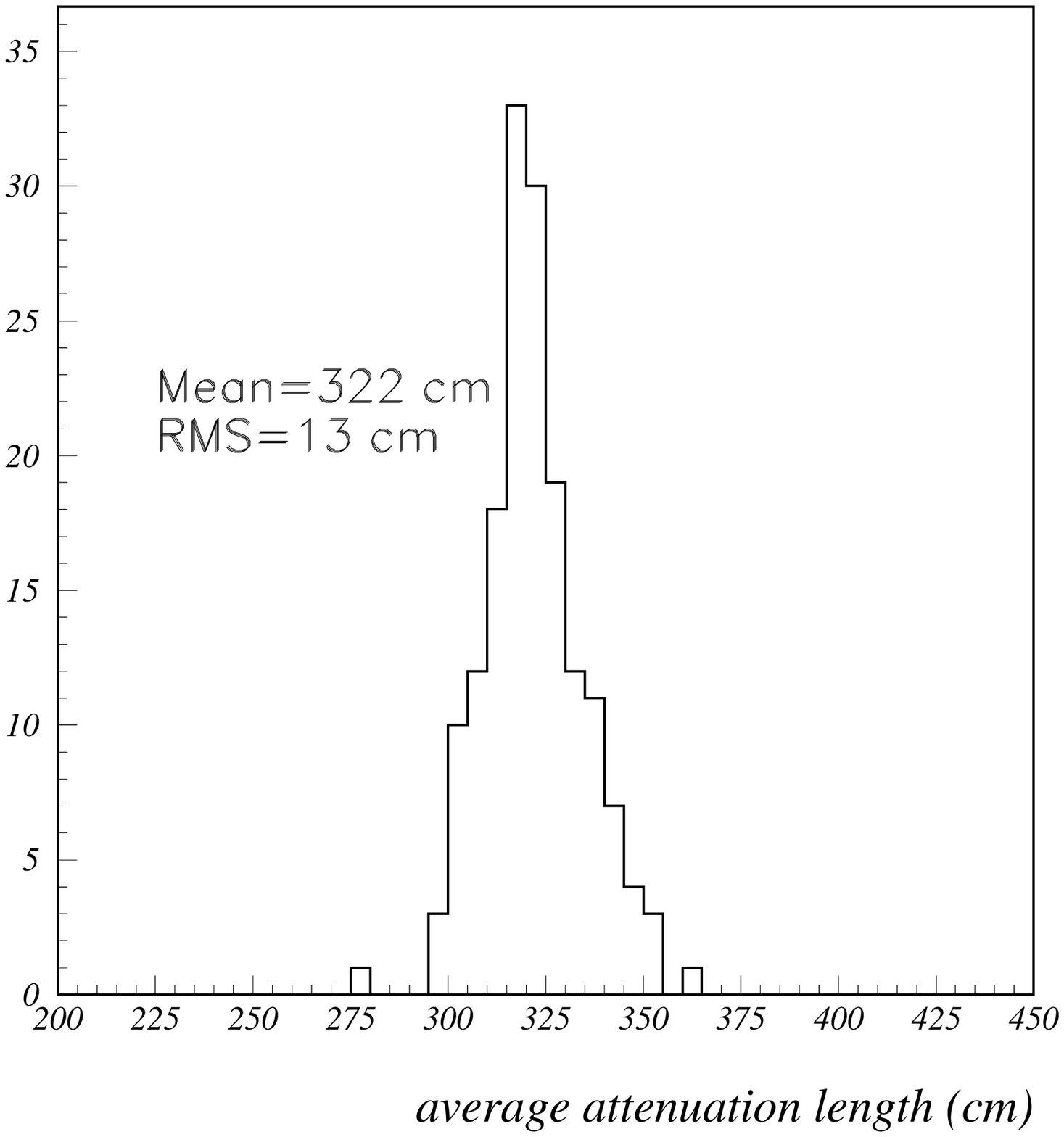}{1.1}{0}}
    \end{center}
  \end{minipage}\hfill
\caption{\sl\it Summary of the average attenuation length for 170 tested preforms.}
\label{fig:summa_at}
\end{figure}

\begin{figure}[e]
 \begin{minipage}{.495\linewidth}
    \begin{center}
      \mbox{\figura{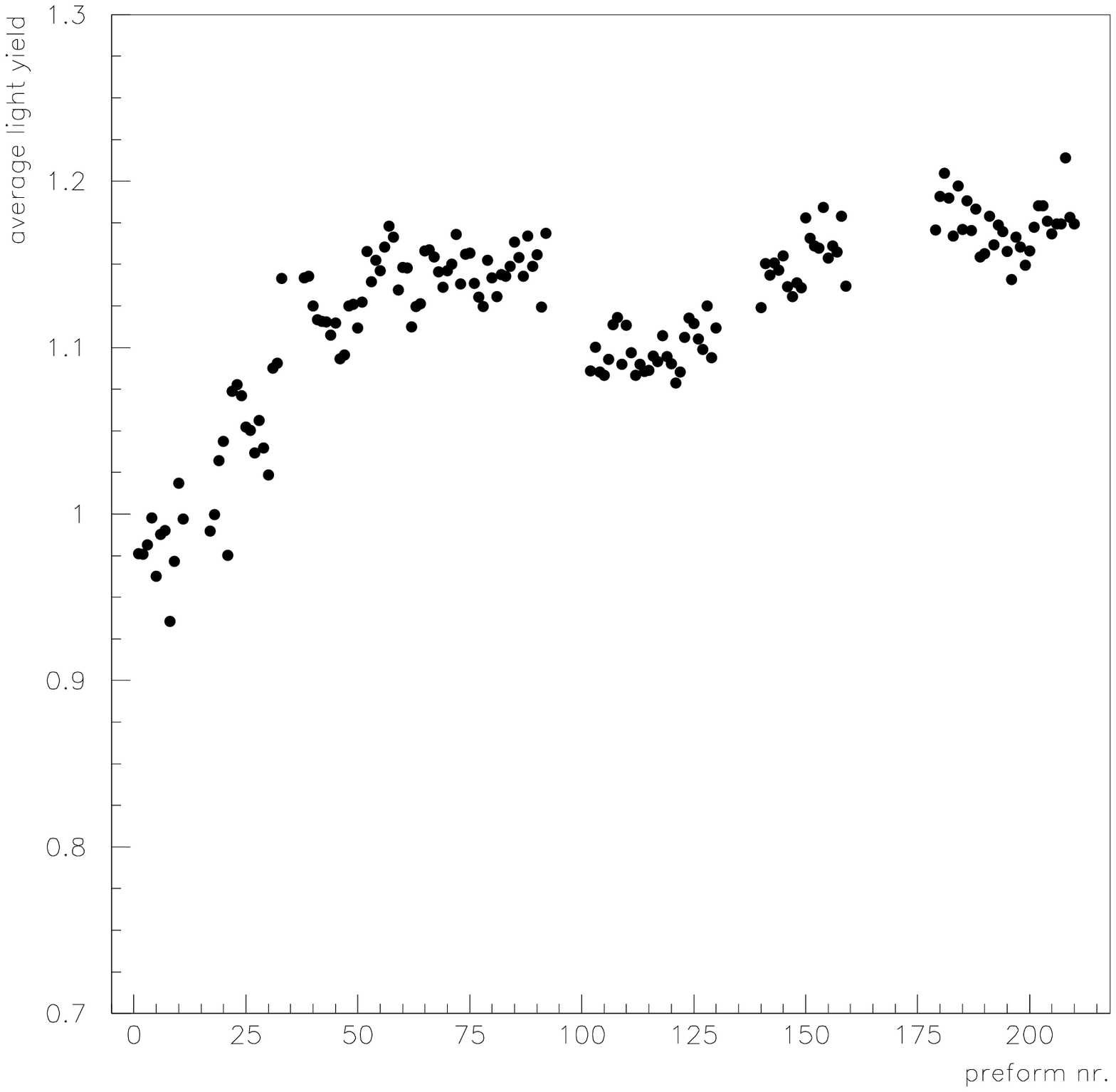}{1.1}{0}}
    \end{center}
  \end{minipage}\hfill
  \begin{minipage}{.495\linewidth}
    \begin{center}
      \mbox{\figura{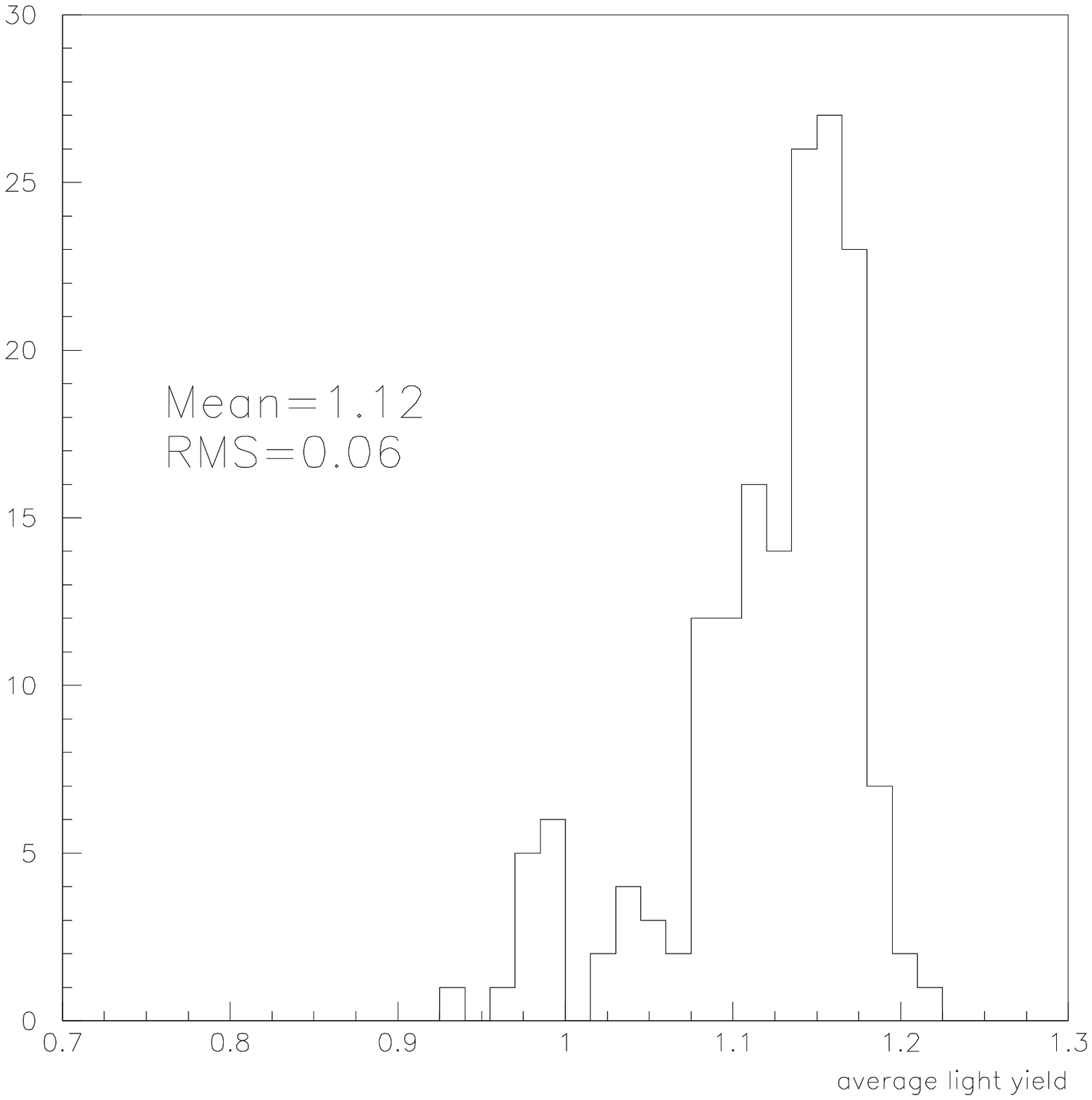}{1.1}{0}}
    \end{center}
  \end{minipage}\hfill
\caption{\sl\it Summary of the average light yield for 170 tested preforms.}
\label{fig:summa_ly}
\end{figure}

From the fibres of each preform, we random sampled 18~fibres
longer than 2~m. The fibres were prepared, as described above i.e.:
\begin{itemize}
\item one end roughly cut and painted with black ink;
\item the other end diamond-cut and polished.
\end{itemize}
The fibres were placed on the plate and coupled to the clear fibres.
Fibres 1 and 11 were the reference fibres and were never removed.
The  scan was performed under computer control and the
subsequent  off line analysis 
provided the results for the acceptance procedure. These are  shown in
 Fig.~\ref{fig:prefesample} for one of the measured preforms.
The ``acceptance sheet''  in the same figure, shows the attenuation length
$\lambda$ and the light yield   $n_0$ for the 18~fibres under test.
Also shown in the lower part of  Fig.~\ref{fig:prefesample} 
 are some important statistical quantities calculated for this preform
together with the acceptance limits of our QC procedure.  
Fig.~\ref{fig:summa_at} shows the attenuation length of all measured preforms
and their distribution.
 Fig.~\ref{fig:summa_ly} shows  results of the same analysis
 for the light yield.
While the attenuation length was rather constant, the light yield
shows an increase in the first part of the production. 
Having  this analysis  been carried out on-line, we have been able
 to notify to the
fibre producer the change  in this parameter, which  they  have than  been able
to correct.

\section{Conclusions}
We have built and put into operation a system to measure in a 
fast and accurate way the optical parameters of  WLS fibres.
In this paper we have described the first version of the set up
that could measure 20 fibres at a time. The system was later improved 
to measure 40 fibres at a time.

The resolutions that we obtain using this device are:    
\begin{itemize}
\item $\Delta n_{0}/n_{0}$=1.5\% for the light yield;
\item $\Delta\lambda /\lambda$=3\% for the attenuation length.
\end{itemize}

This device was used  for the characterization of the fibres that will
instrument the hadron calorimeter of ATLAS (Tile Calorimeter), with excellent 
precision and stability over a long period of measurement.

\section{Acknowledgements}
We thank our technicians R.~Ruberti, F.~Mariani for the support
provided in the construction of the mechanical parts of the apparatus,
and R.~Romboli for the help provided during the fibre characterization.


\begin{thebibliography}{99}

\bibitem{ua2} J. Alitti et al., NIM A263(1988) 51.

\bibitem{d0}  D$\emptyset$ collaboration, ``The D$\emptyset$ Upgrade'' FERMILAB Pub-96/357-E(1996). 

\bibitem{kloe} KLOE collaboration, ``KLOE Detector Tecnical Proposal'',
LNF-93/002 (1993). 

\bibitem{chorus} D. Acosta et al., NIM A308,481 (1991).

\bibitem{h1} H1 Collaboration, ``Tecnical Proposal to Upgrade the
Backward region of the H1  Detector'', PRC 93/02 1993.

\bibitem{zeus} Zeus-Presampler Group, `` A Detector For HERA'',DESY, (1993).

\bibitem{delphi}  DELPHI collaboration, preprint CERN-LEPC/92-6. 

\bibitem{atlas}  ATLAS collaboration, ``Technical Proposal for a
General Purpose pp experiment at the LHC at CERN'', CERN/LHCC/94-43,1994.    

\bibitem{TileCAlTDR} Tile Calorimeter collaboration, `` Atlas Tile Calorimeter 
Technical Design Report'', CERN/LHCC/96-42,1996.

\bibitem{method89} T. Del Prete et al., ATLAS internal
note\footnote{The ATLAS internal notes can be found at
http://weblib.cern.ch in postscript format.}, TILECAL-NO-089(1996).

\bibitem{method99} V. Cavasinni et al., to be submitted as an
 ATLAS internal note.

\bibitem{comment_att} V. Cavasinni et al.,  ATLAS internal note,
TILECAL-NO-004 (2000).
\end{thebibliography}
\end{document}